# Pyrochlore Oxide Superconductor $Cd_2Re_2O_7$ Revisited

Zenji Hiroi,[1*] Jun-ichi Yamaura,[2] Tatsuo C. Kobayashi,[3] Yasuhito Matsubayashi,[1] and Daigorou Hirai[1]

[1]*Institute for Solid State Physics, University of Tokyo, Kashiwa, Chiba 277-8581, Japan*
[2]*Materials Research Center for Element Strategy, Tokyo Institute of Technology, Yokohama, Kanagawa 226-8503, Japan*
[3]*Graduate School of Natural Science and Technology, Okayama University, Okayama 700-8530, Japan*

The superconducting pyrochlore oxide $Cd_2Re_2O_7$ is revisited with a particular emphasis on the sample-quality issue. The compound has drawn attention as the only superconductor ($T_c$ = 1.0 K) that has been found in the family of α-pyrochlore oxides since its discovery in 2001. Moreover, it exhibits two characteristic structural transitions from the cubic pyrochlore structure, with the inversion symmetry broken at the first one at 200 K. Recently, it has attracted increasing attention as a candidate spin-orbit coupled metal (SOCM), in which specific Fermi liquid instability is expected to lead to an odd-parity order with spontaneous inversion-symmetry breaking [L. Fu, Phys. Rev. Lett. **115**, 026401 (2015)] and parity-mixing superconductivity [V. Kozii and L. Fu: Phys. Rev. Lett. **115** (2015) 207002; Y. Wang et al., Phys. Rev. B **93** (2016) 134512]. We review our previous experimental results in comparison with those of other groups in the light of the theoretical prediction of the SOCM, which we consider meaningful and helpful for future progress in understanding this unique compound.

## 1. Introduction

$Cd_2Re_2O_7$ is the only superconductor to be found in the family of α-pyrochlore oxides.[1-3] At room temperature, it crystallizes in a cubic pyrochlore structure of the space group $Fd$–$3m$ with $Re^{5+}$ ions in the $5d^2$ configuration forming a regular corner-sharing tetrahedral network. It exhibits poor metallic behavior at high temperatures, as shown in Fig. 1; this weakly temperature-dependent resistivity continues up to 600 K.[4] Upon cooling, the compound undergoes two transitions at $T_{s1}$ ~ 200 K and $T_{s2}$ ~ 120 K and shows good metallic behavior before superconductivity sets in at $T_c$ ~ 1.0 K. Characteristic structural changes occur at the two transitions possibly by distortion of the Re tetrahedra as schematically depicted in Fig. 1.[5] The three phases are called phases I, II, and III in order of decreasing temperature.

There is increasing interest in its superconductivity[6,7] as it occurs in the low-temperature tetragonal phase, which lacks inversion symmetry in the crystal structure. Such noncentrosymmetric superconductivity is found in $CePt_3Si$,[8] $Li_2Pt_3B$,[9] $CeRhSi_3$,[10] $UIr$,[11] and $CeIrSi_3$,[12] and has attracted much attention, because the absence of inversion symmetry can cause unconventional superconductivity through both spin-singlet and triplet channels in the presence of a strong spin-orbit coupling (SOC).[6,8,13] In addition, unusual vortex states such as a helical vortex phase have been predicted.[14] In this context, $Cd_2Re_2O_7$ is unique because it loses inversion symmetry via the structural phase transition at $T_{s1}$, above $T_c$, which is not the case for other noncentrosymmetric superconductors without inversion symmetry in the whole temperature range. Thus, for $Cd_2Re_2O_7$, one would expect exotic phenomena induced by fluctuations associated with the inversion-symmetry breaking (ISB) transition. Moreover, SOC must be strong in such a $5d$ band semimetal, as indicated by first-principles calculations.[15,16] $Cd_2Re_2O_7$ has another advantage for studying noncentrosymmetric superconductivity: it has a relatively simple normal state without magnetic order or Kondo-like behavior such as found in $CePt_3Si$.[6,8] Hence, it is intriguing to examine the superconductivity of $Cd_2Re_2O_7$ in detail.

The superconductivity of $Cd_2Re_2O_7$ at ambient pressure (AP), however, has been demonstrated to be of the simple $s$-wave, weak-coupling BCS type according to the results of experiments based on heat capacity,[3,17] Re nuclear quadrupole resonance (NQR),[18] Cd NMR,[19] μSR,[20,21] and point-contact spectroscopy.[22] This is possibly because the spin-triplet component can remain minor compared with the spin-singlet component; the ratio between the two components may depend on the pairing mechanism and the magnitude of SOC.[6,13] In contrast, recent high-pressure (HP) resistivity measurements by Kobayashi et al. revealed large changes in the superconducting properties toward the critical pressure of $P_c$ ~ 4.2 GPa: $T_c$ is increased to 2.5 K and the upper critical field $B_{c2}$ becomes 27 times larger than that at AP, which exceeds the Pauli-limiting field.[7] Since such a large $B_{c2}$ has been assumed to be a hallmark of parity-mixing superconductivity in a noncentrosymmetric superconductor,[8] this result strongly indicates that an exotic superconductivity is in fact realized with increasing spin-triplet component under HP in $Cd_2Re_2O_7$.

In addition, the HP study uncovered many electronic phases, at least five phases, near $P_c$ in addition to the three phases present at AP, which indicates certain electronic instability in $Cd_2Re_2O_7$.[7] Very recently, Yamaura et al. investigated the crystal structures under HP and found that the ISB line vanishes at around $P_c$, indicating a close relationship between the ISB and the enhancement of superconductivity.[23] Moreover, they demonstrated that the most electronic phases identified by resistivity are distinguished by their crystal structures, which suggests intimate electron-lattice couplings via strong SOC.

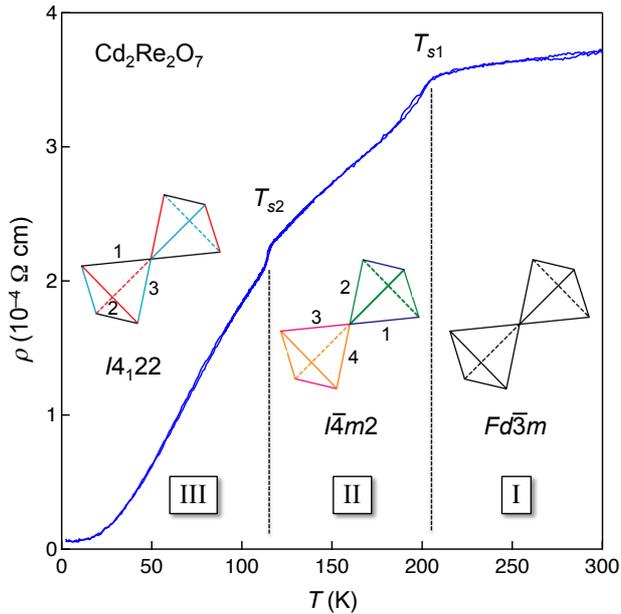

**Fig. 1.** (Color online) Resistivity of $Cd_2Re_2O_7$ from a high-quality crystal (#40A) with RRR ~ 60. Three phases, I, II, and III, appear successively upon cooling through two transitions at $T_{s1}$ = 205 K and $T_{s2}$ = 112 K. The pair of tetrahedra illustrates how the Re tetrahedron network can deform in each phase; the identical bonds are distinguished by the colors and numbers. The two tetrahedra are identical in phase I (space group $Fd$–$3m$), while they become nonequivalent with two kinds of edges for each in phase II ($I$–$4m2$). In phase III ($I4_122$), they again become equivalent with three kinds of edges; there is bond chirality.

In 2015, Fu put forward an intriguing theoretical proposal about spin-orbit-coupled metals (SOCMs).[24] In the spintronics field, antisymmetric spin-orbit coupling (ASOC) in the absence of inversion symmetry has been playing an important role as it can generate sizable spin-split bands that are tunable by the applied field. Fu's proposal is that, even in a metal with inversion symmetry and with strong SOC, a certain Fermi liquid instability leading to the emergence of ASOC is expected, which causes spontaneous ISB to either multipolar, ferroelectric, or gyrotropic order;[24] Norman proposed that these orders can be distinguished by their natural circular dichroism.[25] On the other hand, Kozii and Fu[26] and Wang et al.[27] pointed out the possibility of exotic superconductivity near the quantum critical point (QCP) of the ISB order. They suggested that $Cd_2Re_2O_7$ can be an ideal realization of an SOCM and a suitable target material to study the proposed intriguing physics; very recently, in fact, Harter et al. reported the formation of domains of an odd-parity nematic order below $T_{s1}$ in $Cd_2Re_2O_7$ on the basis of second-harmonic optical anisotropy measurements.[28] Moticated by these fascinating theoretical conjectures, various experiments are now under way, particularly under HP.

The aim of the present paper is to review the previous experimental results on $Cd_2Re_2O_7$ by our group since the discovery of its superconductivity in 2001, particularly in the light of the recent theoretical predictions for SOCMs, and discuss them in comparison with those of the other groups. Because a "new era" has begun since the HP study in 2011 and the theoretical conjectures in 2015–16, it would be helpful to sum up all the previous studies and to share the present status of understanding of the compound. In particular, we would like to address a few controversial issues in the previous reports. Moreover, we will emphasize the importance of the sample-quality issue by including new results of bulk measurements using high-quality crystals obtained recently; the crystals used in previous studies had residual resistivity ratios (RRR; defined as the ratio of the resistivities at 300 and 2 K) of about 10–40, while our recent crystals had larger RRRs of ~60, details of which will be reported elsewhere.[29] We consider it necessary to avoid confusion arising from the sample-quality issue for future progress in understanding the superconducting and ASOC-related phenomena in the present compound representing an SOCM.

This paper is organized as follows. First we will review the crystal and basic electronic structures of $Cd_2Re_2O_7$ in Sects. 2 and 3, respectively. Next, three issues are addressed: the superconducting properties in Sect. 4, the low-temperature (LT) resistivity in Sect. 5, and the two high-temperature (HT) transitions in Sect. 6. Then, after describing the possible orbital or multipolar fluctuations in Sect. 7, we will address the HP properties in Sec. 8, before summarizing the paper in Sect. 9.

## 2. Crystal Structure
### 2.1 Successive changes in the crystal symmetry

The successive changes in the crystal symmetry at the two transitions are summarized here.[5] As illustrated in Fig. 1, two identical Re tetrahedra connected at a corner in phase I become non-equivalent with two kinds of edges for each in $I$–$4m2$ (II) as a result of the missing inversion center; because the threefold axis is also lost at the same time, a tetragonal distortion occurs. Then, in $I4_122$ (III), they again become equivalent by adding a twofold axis and losing the mirror plane at the connecting Re site; now each tetrahedron has three kinds of edges in this chiral space group; note that the Re tetrahedron in $I4_122$ has bond chirality.[5] The relation between these space groups is understood in terms of the translationengleiche subgroups of the mother space group

*Fd*–3*m*: the I–II transition is of the second order as both phases belong to the same branch of symmetry reduction, whereas the II–III transition is of the first order as the two phases are in different branches.[30]

Landau theory based on group-theoretical considerations has been applied to interpret the successive structural phase transition of $Cd_2Re_2O_7$.[31,32] The order parameter involves a doubly degenerate long-wavelength $E_u$ phonon mode at the zone center. Below $T_{s1}$, softening of this mode results in two crystal structures, which may be related to the instability of the Re tetrahedral network.[31] When provided with a certain set of parameters in a thermodynamic potential containing up to sixth-order terms, a series of transitions from *Fd*–3*m* to *I*–4*m*2 and further to $I4_122$ are well reproduced as a function of temperature;[31,32] note that the two LT phases can be perfectly degenerate for a thermodynamic potential without the sixth-order terms.[33] The *I*–4*m*2 phase was suspected to exhibit "metallic ferroelectricity",[34] but it should be a "piezoelectric metal" in which the conductivity can be modified by adjusting the shear stress.[32,35]

A key feature that makes the ISB transition at $T_{s1}$ unique is that the two LT structures are nearly degenerate: the transition can have an unusual tensor character with the order parameter corresponding to the $E_u$ representation.[31,33,34,36] As a result, the low-energy excitation can be a Goldstone phonon mode having a linear dispersion or a long-wavelength fluctuation between the two structures.[33,36] The Raman scattering experiment by Kendziora et al. found such a Raman-active soft mode.[33] On the other hand, very recent second-harmonic optical anisotropy measurements by Harter et al. indicate that the true order parameter is not $E_u$ but $T_{2u}$.[28]

The structural transition at $T_{s1}$ has been clearly evidenced by single-crystal X-ray diffraction (XRD),[5,37,38] powder neutron diffraction,[39] electron diffraction,[40] convergent electron diffraction,[41] Raman spectroscopy,[33,42] and nonlinear optics.[28,36] In contrast, there is some variability in the previous reports on the structural transition at $T_{s2}$. The early single-crystal XRD experiments found significant changes at $T_{s2}$ in the intensity of some fundamental reflections as well as those forbidden for phase I.[5,43] However, little change was observed in the other structural refinements using a single crystal[37] or by powder neutron diffraction.[39] On the other hand, Raman scattering experiments revealed symmetry lowering below $T_{s2}$,[33,44] while nonlinear optics measurements assigned the lowest-temperature structure to *I*–4*m*2, not to $I4_122$.[36] A possible reason for this controversy will be addressed in the present paper.

We consider that the origin of the successive transitions of $Cd_2Re_2O_7$ is related to the Fermi liquid instability of the SOCM. On the other hand, Tachibana et al. pointed out from a comparison with the insulating pyrochlore oxide $Cd_2Nb_2O_7$ that the interplay between the rigid $ReO_6/NbO_6$ octahedron network and the intervening flexible network made of $Cd_2O(2)$ chains is important for the structural transitions;[45] $Cd_2Nb_2O_7$ exhibits three structural transitions within the *Fd*–3*m* subgroups. Possibly related to this, Knee et al. suggested in their Raman scattering study that the $T_{s2}$ transition is driven by ordering of the Cd atoms.[44] Further study is required to clarify the origin of the structural transitions in $Cd_2Re_2O_7$.

*2.2 Structural parameters of phase I*

Most of the pyrochlore oxides with the general formula $A_2B_2O_7$ crystallize in the cubic pyrochlore structure of space group *Fd*–3*m* (No. 227). The A, B, O(1), and O(2) atoms occupy the 16*d*, 16*c*, 48*f*, and 8*b* Wyckoff positions, respectively (origin choice 2); either A or B atoms form the pyrochlore lattice made of corner-sharing tetrahedra. There are only two variable parameters in the structure: the lattice constant $a$ and the atomic coordinate $x$ of the O(1) atom ($x$, 1/8, 1/8). The $x$ value is important as it determines the distortion of the $BO_6$ octahedron, which is regular at $x = 0.3125$ and is slightly compressed along the local [111] direction at larger $x$ values; the site symmetry of the B atom is –3*m*. All the reported $x$ values are larger than 0.3125 for the two types of pyrochlore oxides of $A^{2+}_2B^{5+}_2O_7$ (~0.32) and $A^{3+}_2B^{4+}_2O_7$ (~0.33).[46]

The values of the lattice constant $a$ and the atomic coordinate $x$ of the O(1) atom of the HT phase I of $Cd_2Re_2O_7$ reported at room temperature are ($a$/Å, $x$) = (10.219, 0.3089),[47] (10.2257, 0.3176),[39] (10.226, 0.319),[1,5] and (10.2251, 0.318).[48] They are in good agreement with each other except for those from Donohue's study;[47] their unusually small $x$ value compared with the standard value is suspicious. In fact, Singh et al.[15] obtained $x = 0.316$ by a relaxation of the internal coordinate in local density approximation (LDA) calculations even while keeping $a = 10.219$ Å from Donohue's study.[47] Thus, we take (10.226, 0.318) as the most reliable values.

The $ReO_6$ octahedron is slightly compressed along the local [111] direction at $x = 0.318$; the O–Re–O bond angles are 87.803 and 92.197°, and the Re–O bond length is 1.937 Å for $a = 10.226$ Å. The importance of such a trigonal distortion of the O octahedron in determining the electronic properties has been observed for the related pyrochlore oxide $Cd_2Os_2O_7$ with $x = 0.319$.[49,50] In this compound, the small trigonal distortion causes large easy-axis anisotropy on the localized Os spins and stabilizes the all-in–all-out-type magnetic order below the metal-insulator transition temperature of 227 K.[49,51]

*2.3 Structural parameters of phases II and III*

The structural parameters of the two LT phases are not yet available or seem to still be unreliable. One of the reasons is that the number of parameters to be refined increases enormously compared with the HT cubic structure because of the symmetry lowering. Moreover, in XRD experiments, the presence of the two heavy elements makes it difficult to obtain reliable values for the atomic positions of the light oxygen atoms, while in neutron diffraction experiments, neutrons rarely escape from a sample because natural Cd (containing 12% $^{113}$Cd) absorbs most of them.

Castellan et al. performed a single-crystal XRD experiment and found a tiny tetragonal distortion of 0.05% at 15 K.[37] The structural parameters of phases II at 160 K and III at 90 K were reported by Huang et al. based on single-crystal XRD experiments.[38] However, we cannot judge their reliability because no details of the experiments were given; the purpose of their paper is on band structure calculations. On the other hand, the powder neutron diffraction results by Weller et al. using a $^{114}$Cd-enriched powder sample may be more reliable but, unfortunately, they were obtained by analyzing a diffraction data taken at 13 K assuming space group $I\text{–}4m2$, not $I4_122$ for phase III.[39]

## 3. Basic Electronic Structure

The electronic structure of phase I has been calculated by three groups. Among them, the structures calculated by Singh et al.[15] and Harima[16] are similar, while that calculated by Huang et al. seems considerably different.[38] According to the former two studies, the compound is a compensated semimetal with nominally equal numbers of holes and electrons in the Re 5$d$ ($t_{2g}$) bands; the estimated carrier density is $1.47 \times 10^{20}$ cm$^{-3}$.[16,52] There are two electron (one hole) Fermi surfaces centered at the $\Gamma$ (K) point. Reflecting the presence of flat bands just below the Fermi level, there are several sharp spikes in the density of states (DOS), as shown in Fig. 2. In contrast, Huang et al. claimed that $Cd_2Re_2O_7$ is an uncompensated semimetal with larger Fermi surfaces.[38]

It is crucial to know the dependence of how the Fermi surfaces of phase I split on the spin component by the ASOC as a result of ISB at $T_{s1}$ and also how the splitting is modified at $T_{s2}$. Huang et al. reported large changes in the Fermi surfaces on the basis of their structural study.[38] In particular, for phase III, they found quasi-two-dimensional electron Fermi surfaces with hole surfaces missing, which was unexpected; a replication study is required.

Harima considered a possible change in the Fermi surfaces induced by the lack of inversion symmetry.[16] As a result of lifting the spin-degeneracy (parity violation splitting), the two electron surfaces split into four; they should survive because the degeneracy at the $\Gamma$ point remains. In contrast, one of the two split hole surfaces should disappear when a small spin splitting of 68 meV occurs at the K point. This can explain the observed large decrease in the DOS at $T_{s1}$;[52] there are 12 hole pockets at the zone boundary and also the hole band may be heavier than the electron band.[15] Moreover, this is qualitatively consistent with the increase in the carrier density below $T_{s1}$.

Concerning the ASOC in the LT phases of $Cd_2Re_2O_7$, one would expect Dresselhaus-type spin splitting[53] because the $T_{s1}$ transition is basically a transition from the diamond to the zinc-blende type. Generally, the inversion-asymmetry splitting is small, but it can be enhanced in the 5$d$ bands of $Cd_2Re_2O_7$, as in HgSe.[54] Note that Harter et al. gave an example of spin-split bands based on the $T_{2u}$ representation.[28] For further quantitative discussion, detailed analyses based on first-principles calculations that use complete and reliable structural parameters are required. Moreover, future experiments that give direct information on the Fermi surfaces will be crucial.

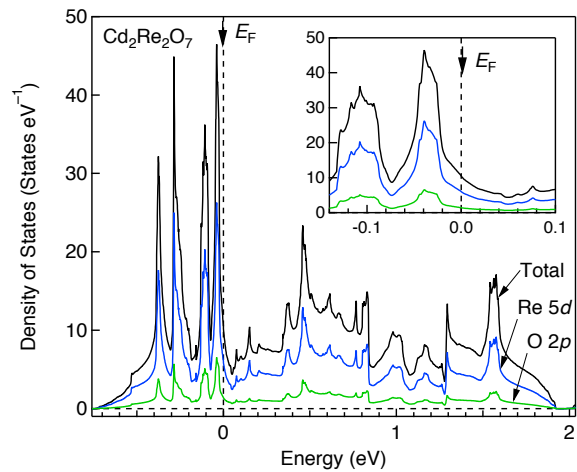

**Fig. 2.** (Color online) Density of states (DOS) of phase I calculated by Harima. The top black curve shows the total DOS, and the blue and green ones below represent the partial DOSs of the Re 5$d$ and O(1) 2$p$ states, respectively. Compared with the previous calculations,[16] the profile is slightly modified by taking into account the relativistic effects on the mass correction in the spin-orbit coupling.[55]

## 4. Superconductivity

In this section, we discuss the superconductivity of $Cd_2Re_2O_7$. We mention the sample dependences of $T_c$ in terms of heat capacity and resistivity in the previous study as well as for recent crystals and discuss them in terms of disorder effects. Then, we analyze the heat capacity data of the best crystal using the $\alpha$ model, which shows that the superconducting gap is not a simple $s$-wave gap but may be slightly modified by the parity-mixing effect for a noncentrosymmetric superconductor.

### 4.1 Heat capacity

There was scatter in the superconducting transition temperature $T_c$ and the upper critical field $B_{c2}$ reported in the previous studies. Hanawa et al. reported $T_c = 1.0$ K from their heat capacity ($C_p$) measurements,[1] while Sakai et al. claimed a slightly higher $T_c$ of 1.1 K from their resistivity ($\rho$) and ac-susceptibility ($\chi_{ac}$) measurements.[2] Later, they obtained $T_c$ = 1.4 K from the $\rho$ data.[56] On the other hand, Jin et al. obtained $T_c$ = 1.15 K from $\rho$ and $\chi_{ac}$ data and $T_c$ = 1.47 K from the onset of a resistive transition; in the same paper they reported $T_c$ = 0.99 K from the $C_p$ data.[3] Furthermore, Vyaselev et al. found a coherence peak at $T_c$ = 0.98 K in their Re NQR experiments.[18] More recently, Haraguchi et al. obtained $T_c$ = 1.75 K from $\rho$, $C_p$, and magnetization measurements on polycrystalline samples.[57] Thus, the reported $T_c$ values are in fact scattered in a considerably large $T$ range.

This is also the case for $B_{c2}$. Hanawa and coworkers reported $B_{c2}$ = 0.21 or 0.29 T from their $C_p$ data,[1,17] while Sakai et al. and Jin et al. obtained larger values of 0.7–0.8 T from $\rho$ under magnetic fields.[2,3] Sakai et al. also reported that $B_{c2}$ < 0.37 T from the Knight shift of Cd NMR.[2] On the other hand, Haraguchi et al. reported $B_{c2}$ ~ 4 T from $C_p$ measurements.[57] We speculate that such scattering in $T_c$ and $B_{c2}$ is related to the measurement methods as well as nonstoichiometry, disorder, or strain in the samples.

Figure 3(a) shows a set of heat capacity data for one polycrystalline sample and four single crystals with RRRs of 20–30 from our previous study[17,58] and also one recent high-quality crystal with RRR ~ 60.[29] The polycrystalline sample shows a broad transition with $T_c$ = 1.14 K at the midpoint of the transition, while the transitions of the single crystals are sharper with $T_c$ = 0.97–1.04 K. Note that there is a clear trend: the sharper the transition, the lower the $T_c$. Among them, crystal #1E (a crystal from the number 1 batch) has the lowest $T_c$ of 0.97 K and the smallest transition width of $\Delta T_c$ = 0.05 K, which are comparable to those reported previously by Hanawa et al.[1] and Jin et al.[3] from their heat capacity measurements. The recent crystal #41A shows an almost identical sharp transition to that of crystal #1E. In contrast, crystal #5C shows a broader transition with $T_c$ = 1.04 K and $\Delta T_c$ ~ 0.2 K; there is a tail on the high-temperature side. Such a broad transition should originate from a distribution of $T_c$ in a crystal, which may be caused by some chemical or physical disorder. Thus, disorder must enhance $T_c$ in $Cd_2Re_2O_7$. The data of crystal #1E will be analyzed in detail in Sect. 4.4.

*4.2 Resistivity*

Resistivity, in contrast to heat capacity, is not always a bulk quantity: a conducting path between two electrodes with the highest $T_c$ governs the variation at the transition. Figure 3(b) shows resistive transitions for seven crystals: crystals #1E, #2A, #7C, and #10A were polished and then chemically etched in dilute hydrochloric acid, while the other crystals were pristine or only polished; crystal #10A has RRR ~ 40 and was used in the HP resistivity study in 2011.[7] The former four crystals have $T_{c0}$ = 0.99, 1.09, 1.02, and 0.97 K, respectively, where $T_{c0}$ is defined as the offset temperature with $\rho$ = 0. Each $T_{c0}$ value is in good agreement with the value of $T_c$ from the corresponding heat capacity data, indicating that the resistive transition also has bulk character. Note that $T_{c0}$ tends to increase slightly with increasing $\rho_0$. Since $\rho_0$ can be a measure of electron scattering induced by disorder, this means again that $T_c$ is enhanced with increasing disorder, as observed in the heat capacity data.

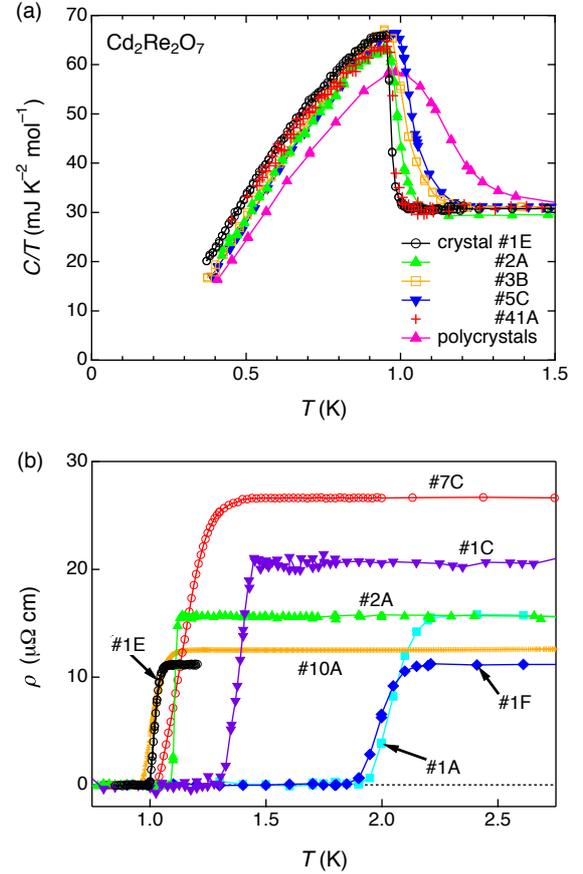

**Fig. 3.** (Color online) Sample dependences of the superconducting transition in terms of the heat capacity (a) and resistivity (b).[17] One polycrystalline and five single-crystal samples including one from this study (#41A) were used for heat capacity measurements. Three kinds of specimen were used for the resistivity measurements in (b): a pristine crystal (#1C), rectangular-bar-shaped crystals after polishing (#1A, #1F), and rectangular-bar-shaped crystals after chemical etching in a dilute HCl solution (#1E, #2A, #7C, #10A).

The $T_c$ values of the pristine crystal #1C and polished crystals #1A and #1F without etching are much higher, approaching to 2 K; $T_{c0}$ = 1.90, 1.30, and 1.85 K, respectively. Since these higher values of $T_c$ have never been observed in our heat capacity measurements, they are not bulk $T_c$ values but must originate from the pristine or polished surface. In our previous experiments, it was clearly demonstrated that, after polishing, $T_{c0}$ from $\rho$ was increased from 1 to 2 K only on the polished surface, with $T_c$ from $C_p$ remaining at 1 K.[58] Moreover, we observed that the etching in dilute hydrochloric acid recovered $T_{c0}$ to ~1.0 K. Although it is difficult to know the phenomena occurring on the polished surface, degradation such as structural defects, chemical nonstoichiometry, or strain may have been introduced, which should increase $T_{c0}$ only near the surface; the etching may remove the degraded surface layer. Therefore, we conclude that the controversial issue on $T_c$ originated from disorder effects in the bulk and, more seriously, at the surface. The large $B_{c2}$ values observed on the basis of resistivity measurements should have the same origin. The intrinsic $T_c$ and $B_{c2}$ for $Cd_2Re_2O_7$ in the clean limit are 0.97 K and 0.2–0.3 T, respectively.[58]

*4.3 Disorder effects*

Let us discuss the effects of disorder on $T_c$, which have been studied thus far on various phonon-mediated superconductors. In the A-15 compounds and bcc alloys with relatively high $T_c$, $T_c$ decreases with increasing $\rho_0$ (disorder), whereas those with low $T_c$ exhibit the inverse trend;[59] for example, as $\rho_0$ increases, $T_c$ decreases from 17.5 to ~2 K in $Mo_3Ge$, whereas it increases from 1.16 to 2.5 K in Al.[60] These facts can be explained by the band-smearing effect:[61] when the DOS has a sharp peak (a steep valley) near the Fermi energy $E_F$, the DOS at $E_F$ may be decreased (increased) as the profile is broadened by disorder. Note that even a small change in the DOS can seriously affect $T_c$ in the conventional BCS formalism of $T_c$. There is also alternative discussion about the effects of disorder on $T_c$ that more generally assumes competition between attractive electron-phonon (*e-ph*) and repulsive electron-electron (*e-e*) interactions: disorder may selectively enhance the former to increase $T_c$ for weak-coupling, low-$T_c$ superconductors, while it enhances the latter to decrease $T_c$ for strong-coupling, high-$T_c$ superconductors.[59]

The observed enhancement of $T_c$ in $Cd_2Re_2O_7$ appears to be in line with that of previous superconductors: $T_c$ decreases with increasing disorder as it is a weak-coupling, low-$T_c$ superconductor. In fact, there is a steep valley at $E_F$ in the DOS,[15,16] as reproduced in Fig. 2. Thus, the band-smearing effect gives a reasonable explanation: a rise in the DOS at $E_F$ should occur with increasing disorder, resulting in a higher $T_c$. On the other hand, note that the disorder effect in $Cd_2Re_2O_7$ is relatively weak compared with those in the *p*-wave superconductor $Sr_2RuO_4$,[62] in which the superconductivity occurs only in samples with $\rho_0$ smaller than 1 $\mu\Omega$ cm.[63] This is consistent with the dominant *s*-wave superconductivity of $Cd_2Re_2O_7$.

Concerning the 2 K transition on the crystal surface, we speculate that a similar band-smearing effect occurs more efficiently because additional disorder may be introduced by polishing. It may also be possible to assume chemical nonstoichiometry, which would change the Fermi level so as to raise the DOS: for example, Cd deficiency would cause hole doping to raise the DOS. Alternatively, strain induced by polishing can cause the enhancement of $T_c$, judging from the fact that the application of pressure increases $T_c$ up to 2.7 K.[4,64,65] A similar enhancement of $T_c$ from 1.5 to 3 K has been observed at an interface in $Sr_2RuO_4$, which may be due to uniaxial pressure.[66] This surface issue regarding $T_c$ for $Cd_2Re_2O_7$ may be related to another controversial issue regarding the $T_{s2}$ transition, as will be mentioned in Sect. 6.

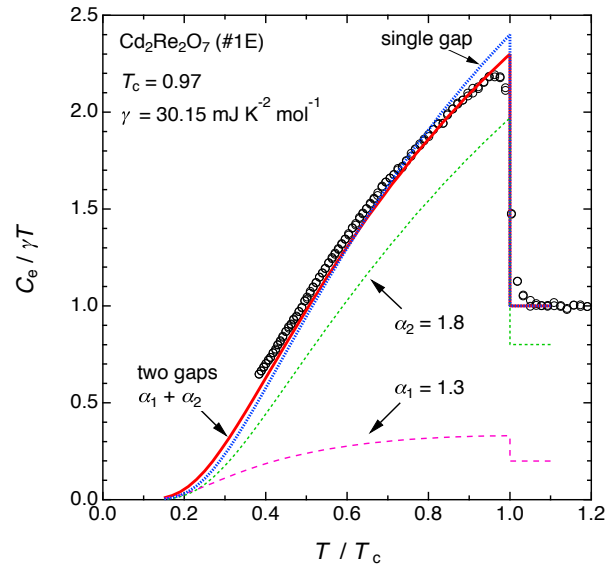

**Fig. 4.** (Color online) Heat capacity from the highest-quality crystal (#1E) fitted to the α model. Dotted (blue) and solid (red) lines near the data marks are fits to single-gap and two-gap models ($\alpha_1 = 1.3$, $\alpha_2 = 1.8$, and $f = 0.2$), respectively. The two separate contributions of the latter fit are shown by the two lower broken lines.

*4.4. Superconducting gap*

We have analyzed in more detail the heat capacity data of the highest-quality crystal #1E. As shown in Fig. 4, the magnitude of the jump at $T_c$ divided by the Sommerfeld coefficient $\gamma$ (30.15 mJ K$^{-2}$ mol$^{-1}$), $\Delta C_e/\gamma T_c$, is 1.15, considerably smaller than 1.43 from the weak-coupling BCS theory; it is almost independent of disorder, as shown in Fig. 3(a). Thus, the simplest BCS fit is clearly not adequate.

We have fitted the data assuming the conventional α model for a multigap superconductor, which has been proved to well reproduce heat capacity data for many superconductors such as $MgB_2$,[67] $Nb_3Sn$,[68] and β-pyrochlore oxides.[69] It is assumed in the α model that two bands with different superconducting gaps, $\Delta_1$ and $\Delta_2$, contribute independently to the heat capacity. Each band is characterized by a partial Sommerfeld constant $\gamma_i$, so that the total $\gamma$ equals $\gamma_1 + \gamma_2$. Heat capacity data are fitted with three parameters: $\alpha_i = \Delta_i/k_B T_c$ ($i = 1, 2$) and $f = \gamma_1/\gamma$. Although the model assumes two gaps, it has been successfully applied to superconductors with an anisotropic gap.[67]

We obtain $\alpha_1 = 1.3$, $\alpha_2 = 1.8$, and $f = 0.2$ for $Cd_2Re_2O_7$, which give a slightly improved fitting compared with a single-gap fitting (Fig. 4). However, the fitting is not still perfect: the data in the intermediate-temperature range around $T_c/2$ are significantly larger than calculated. This suggests that the assumption of two gaps or an anisotropic gap is not appropriate, and a more elaborate picture is required. It would be interesting if this fact reflects a possible effect of parity mixing in the superconducting state, as expected for a noncentrosymmetric superconductor, even if it is small. The magnitudes of the superconducting gap determined by other experiments are $\Delta/k_B T_c$ = 1.84 by Re NQR,[18] 2.3 by Cd NMR,[19] and 2.55 by point-contact spectroscopy.[22] The origin of this scattering is not known but may be related to the complex form of the superconducting gap of $Cd_2Re_2O_7$.

Finally, we note two recent observations on the superconductivity of $Cd_2Re_2O_7$. One is a point-contact-spectroscopy study by Razavi et al., which suggests the presence of a phase transition at ~0.8 K within the superconducting region.[22] The other is from optical reflectance measurements by Hajialamdari et al., which found the emergence of two low-energy absorption peaks at 1.2 and 2.4 meV at 0.5 K only in the superconducting state.[70] These surprising results remain to be confirmed but may reflect an exotic aspect of the superconductivity of $Cd_2Re_2O_7$.

## 5. Low-Temperature Resistivity

Next we consider the temperature dependence of the resistivity of $Cd_2Re_2O_7$ at low temperatures. In general, it provides important information on low-energy excitations that govern the scattering of carriers.[71] It is well known that low-temperature resistivity takes a $T^5$ form for simple, long-wavelength phonon scattering, as expected from the Bloch–Grüneisen relation, while it takes a $T^2$ form for dominant $e$-$e$ scattering in a Fermi liquid or strong $e$-$ph$ scattering in specific materials: bismuth with a small cylindrical Fermi surface,[72] A-15 compounds with extremely strong $e$-$ph$ couplings,[73] and β-pyrochlore oxides with a large electron-rattler (anharmonic oscillator) coupling.[69,74,75] Moreover, $T^2$ resistivity at very low temperatures can arise from inelastic scattering by the thermal motion of nonmagnetic impurity ions, as predicted by Koshino[76] and Taylor[77] and actually observed in K-Rb alloys.[78] On the other hand, $T^3$ dependence has been observed in various elements and compounds but, in most cases, it is fortuitous and transitional between the $T^5$ and $T^2$ regimes at low and high temperatures, respectively.[79] In fact, many transition metals eventually show $T^2$ behavior at very low temperatures after a $T^3$ dependence.[71] Therefore, the $T^3$ dependence is unusual. On the other hand, $T^n$ behavior with $n$ smaller than 2 occurs in compounds with magnetic fluctuations such as heavy-fermion compounds, but this is apparently not the case for $Cd_2Re_2O_7$ without any tendency of magnetic ordering.

*5.1 $T^2$ resistivity of $Cd_2Re_2O_7$*

In previous studies, Jin et al. and Huo et al. reported resistivity proportional to $T^2$ with large coefficients of $A$ = 0.024 and 0.012 μΩ cm K$^{-2}$ in wide $T$ ranges below 60 and 20 K, respectively.[3,56] This $T^2$ resistivity has been considered as evidence for strong electron correlations dominating the electron transport and cited in later studies.[4,28,31,35] In contrast, Hanawa et al.[1] and Barišić et al.[4] reported a $T^3$ dependence at lower temperatures. In order to clarify this discrepancy, we have carefully examined the sample dependence of the LT resistivity using high-quality crystals with RRR = 30–60.

Figure 5(a) shows $\rho$ as a function of $T^2$ for eight selected crystals with various RRR and $\rho_0$ values. The $T^2$ plot is concave upward below 30 K and shows approximately linear dependence below ~15 K. The coefficients of the $T^2$ term $A$ ($A_2$) and the residual resistivity $\rho_0$ for all 16 crystals examined are 0.006–0.06 μΩ cm K$^{-2}$ and $\rho_0$ = 5–70 μΩ cm, respectively, compared with those of 0.024 μΩ cm K$^{-2}$ and $\rho_0$ = 17 μΩ cm below 60 K reported by Jin et al.[3]

In the absence of extremely strong $e$-$ph$ couplings, as is probably the case for $Cd_2Re_2O_7$, the $T^2$ term can be ascribed to electron-electron scattering and inelastic impurity scattering: $A$ is made up of two parts, $A = A_0 + A_I\rho_0$, where $A_0$ and $A_I$ are the contributions from them, respectively.[78] Figure 5(b) plots $A$ for the $Cd_2Re_2O_7$ crystals as a function of $\rho_0$. There is a clear tendency that $A$ becomes large with increasing $\rho_0$. A linear fit to the above equation gives $A_0$ = 9(4) × 10$^{-9}$ Ω cm K$^{-2}$ and $A_I$ = 5.8(8) × 10$^{-4}$ K$^{-2}$. Thus, the $T^2$ term mainly originates from the Koshino–Taylor term, and the intrinsic term is small; $A_0$ is much larger than 2.2 × 10$^{-13}$ Ω cm K$^{-2}$ for K-Rb alloys[78] but much smaller than those for conventional strongly correlated electron systems of 10$^{-9}$–10$^{-4}$ Ω cm K$^{-2}$.[80]

Note that $T^2$ resistivity manifests itself under high pressures.[7,64] As the pressure increases, a $T^2$ dependence emerges with the coefficient $A$ increasing from a small value of 0.004 μΩ cm K$^{-2}$ at AP to 0.01 μΩ cm K$^{-2}$ at 3 GPa.[7] Then, $A$ rapidly increases to 0.05 μΩ cm K$^{-2}$ at 4.2 GPa, where the ISB transition line vanishes.[23] Since it is unlikely that electron correlations are enhanced by pressure, this $T^2$ resistivity is not related to electron correlations but to a fluctuation associated with the ISB transition. Conversely, this fact suggests that the origin of the intrinsic $A_0T^2$ term at AP is due to the same fluctuation of a smaller magnitude or to weak electron correlations, as discussed above.

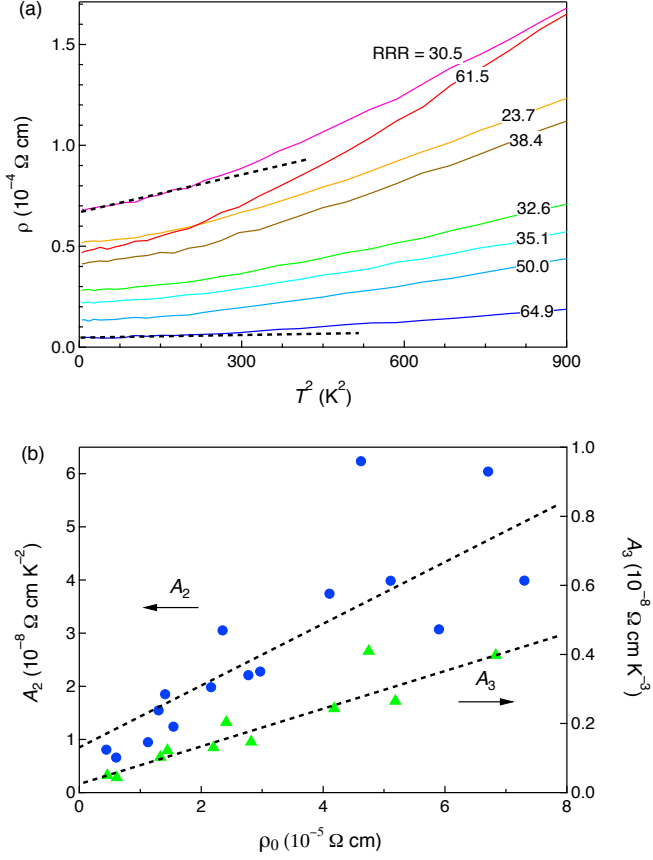

**Fig. 5.** (Color online) (a) Resistivity data below 30 K plotted against $T^2$ for various crystals with different RRRs. The broken lines show linear fits below 15 K. (b) Coefficients of the $T^2$ ($A_2$) and $T^3$ terms ($A_3$) plotted against the residual resistivity $\rho_0$. The broken lines are linear fits.

### 5.2. $T^3$ dependence of $\rho$

We have performed precise resistivity measurements on a high-quality crystal (#10A) with RRR ~ 40, the results of which are plotted as functions of $T$, $T^2$, and $T^3$ below 30 K in Fig. 6. The $\rho$-$T$ curve is almost flat below 10 K, far below the Debye temperature of 458 K,[1] and the $\rho$-$T^2$ curve shows linear dependence below ~8 K with $A = 0.00605(5)$ $\mu\Omega$ cm K$^{-2}$. In contrast, the $T^3$ plot becomes linear asymptotically below ~17 K. Alternatively, we have fitted the data below 15 K to the form $\rho = \rho_0 + A'T^n$ and obtained $\rho_0 = 12.594(1)$ $\mu\Omega$ cm, $A' = 0.0042(1)$ $\mu\Omega$ cm K$^{-n}$, and $n = 3.17(1)$. Thus, the LT resistivity is well reproduced by a $T^3$ term.

Similarly to the $T^2$ term, the coefficient $A_3$ of the $T^3$ term depends on $\rho_0$ and tends to become small in the limit of $\rho_0 = 0$, as shown in Fig. 5(b); the residual $A_3$ is $2.5 \times 10^{-10}$ $\Omega$ cm K$^{-3}$. Thus, the $T^3$ term may not be intrinsic. In fact, the $\rho$ data below 15 K can be reasonably reproduced by the form $\rho = \rho_0 + A_2T^2 + A_5T^5$ with $\rho_0 = 12.491(2)$ $\mu\Omega$ cm, $A_2 = 6.63(4) \times 10^{-9}$ $\Omega$ cm K$^{-2}$, and $A_5 = 1.011(7) \times 10^{-12}$ $\Omega$ cm K$^{-5}$, as shown for the $\rho$–$T$ curve in Fig. 6, which suggests that the $T^3$ dependence is fortuitous and transitional between the two conventional terms.[79]

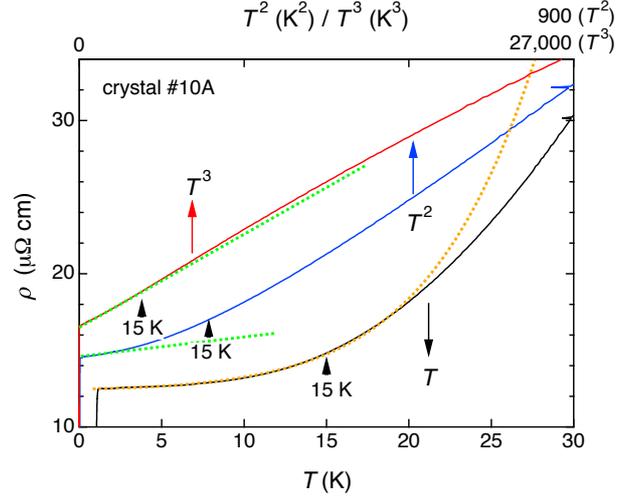

**Fig. 6.** (Color online) Resistivity of crystal #10A below 30 K as functions of $T$ (bottom), $T^2$ (middle), and $T^3$ (top). The $T^2$ and $T^3$ plots are shifted upward by 2 and 4 $\mu\Omega$ cm, respectively, for clarity. The (green) broken lines on the $T^2$ and $T^3$ plots are linear fits, and the (orange) broken curve on the $T$ plot is a fit to the form $\rho = \rho_0 + A_2T^2 + A_5T^5$.

### 5.3. Electron correlations?

Summarizing this section, the $T^2$ contribution to $\rho$ is small and mostly originates from the Koshino–Taylor term. Thus, electron correlations must play a minor role in Cd$_2$Re$_2$O$_7$. On the other hand, the importance of electron correlations was inferred from the $T$ dependence of the magnetic susceptibility $\chi$ at high temperatures of above 400 K, which seems to follow the Curie–Weiss law.[43,81] However, the estimated Weiss temperatures are unusually high, ~1,500 K, in spite of the absence of antiferromagnetic correlations according to an NMR study;[18] instead, a ferromagnetic correlation is suggested. The $T$ dependence of $\chi$ may originate from the complex band structure of Cd$_2$Re$_2$O$_7$.[15,16] On the other hand, a muon spin rotation spectroscopy experiment appears to provide further evidence of electron correlations:[82] the presence of a spin polaron, which is a localized state via strong exchange interactions in a correlated metal, was suggested.

## 6. $T_{s1}$ and $T_{s2}$ Transitions

The $T_{s1}$ transition has a marked effect on the crystal structure and various physical quantities and is always obvious in any sample. In contrast, the second transition at $T_{s2}$ remains somewhat unclear. Anomalies in various quantities observed at $T_{s2}$ are much weaker or sometimes indiscernible. Here we summarize previous observations on the two transitions and discuss the source of the controversy on the $T_{s2}$ transition on the basis of bulk measurements using two crystals with RRR ~ 60.

### 6.1 Changes in the electronic properties at $T_{s1}$

The magnetic susceptibility,[1,40,48,81,83] $^{111}$Cd NMR Knight shift, and spin-lattice relaxation rate[18,81,84] exhibit distinct anomalies at $T_{s1}$ but almost no changes at $T_{s2}$, as shown in Fig. 7. Since the Re NQR spectrum shows sharp peaks in the whole $T$ range, there is neither magnetic order nor a nonuniform charge distribution of $5d$ electrons.[18] Thus, the observed decreases at $T_{s1}$ must be due to a reduction in the DOS. Note that the DOS becomes nearly half at the lowest temperature. On the other hand, the large decrease in resistivity below $T_{s1}$ is possibly due to the increase in the carrier density. On the basis of these experimental findings, the mechanism of the $T_{s1}$ transition has been assumed to be a band-Jahn–Teller transition.[16,52,83]

In the light of the recent theoretical proposal of the SOCM,[24] the $T_{s1}$ transition may be caused by the specific Fermi liquid instability of the SOCM. The ISB spontaneously occurs so as to induce spin splitting on the Fermi surface of phase I. The heavy hole bands at the zone boundary probably spin-split owing to the ASOC activated by the ISB and partially disappear below the Fermi level. Phase II may be a multipolar phase, which should be evidenced in future experiments.

### 6.2 $T_{s2}$ transition

The presence of the $T_{s2}$ transition has been noted on the basis of resistivity,[83] heat capacity,[45,83] magnetic susceptibility,[45] Re NQR,[85] thermoelectric power,[4,56,57,86] magnetoresistance,[83] and Hall coefficient measurements.[86] For example, a clear anomaly in resistivity with the thermal hysteresis characteristic of a first-order transition has been observed in a crystal, whereas only inflections or smooth changes have been observed for many other crystals.[4,40,45,48,56,86-88] The anomalies in the heat capacity are small compared with those at $T_{s1}$ but give direct evidence of a thermodynamic phase transition.[45,83] The estimated entropy changes $\Delta S$ are 0.08–0.10 J K$^{-1}$ mol$^{-1}$, which are much smaller than that of the $T_{s1}$ transition (7.0 or 3.8 J K$^{-1}$ mol$^{-1}$).[40,83] Thus, the changes in the free energy associated with the structural and electronic changes must be small at $T_{s2}$. In contrast, there are reports that claim the absence of heat capacity anomalies.[40,87] In addition, it was suggested from the results of photoemission spectroscopy and scanning tunneling microscopy, both of which are surface-sensitive, that the $T_{s2}$ transition is not an intrinsic phase transition but is driven by certain imperfections.[87] We will address a possible reason for this controversy in the next section.

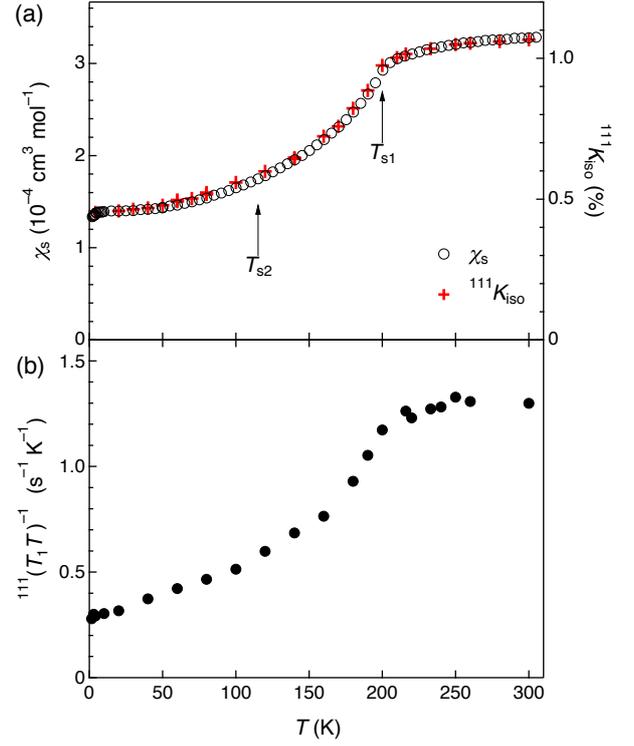

**Fig. 7.** (Color online) (a) Spin susceptibility $\chi_s$ (left) and isotropic component of Knight shift $^{111}K_{iso}$ (right) from $^{111}$Cd NMR experiments.[18] They decrease gradually below $T_{s1}$ with no anomalies at $T_{s2}$. (b) $^{111}(T_1T)^{-1}$ from the $^{111}$Cd NMR experiments shows a similar decrease below $T_{s1}$ and no anomaly at $T_{s2}$, while it continues to decrease below 50 K in contrast to $\chi_s$ and $^{111}K_{iso}$.

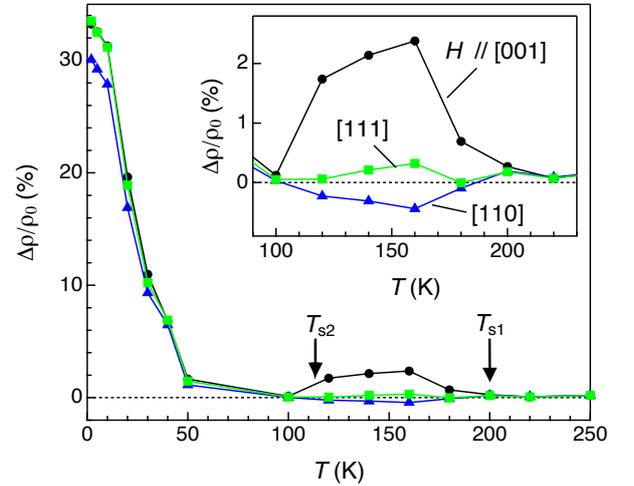

**Fig. 8.** (Color online) Magnetoresistances measured at a magnetic field of 14 T applied along the [001], [110], and [111] directions with an electrical current along the [1–10] direction of the pseudo-cubic structure.[83]

The magnetoresistance data reproduced in Fig. 8 is extremely interesting and strongly indicates what happens to the electronic structure at $T_{s2}$.[83] The almost isotropic, positive $\Delta \rho_m$ observed at 2 K decreases upon heating and vanishes at 100 K. Then, anisotropic $\Delta \rho_m$ appears in an intermediate-$T$ window before vanishing again above 200 K. Note that a negative magnetoresistance is observed at $B$ parallel to the cubic $[110]_c$ direction. Thus, only phase II possesses an anisotropic magnetoresistance, indicating a specific Fermi surface, and the transition to phase III at $T_{s2}$ must be accompanied by a significant change in the Fermi surface.

The $T_{s2}$ transition is clearly detected by thermopower measurements,[4,56,57,86] which are usually sensitive to a change in the Fermi surface. Barišić et al. observed an S-shaped anomaly at 120–130 K as a smoothed-out discontinuity due to the first-order transition.[4] A microscopic probe of Re NQR also found a large discontinuous change in the spectrum at $T_{s2}$;[85] two-phase coexistence was observed at 120 K. All these previous experimental results demonstrate that not a large but a small change takes place on the Fermi surface at $T_{s2}$.

### 6.3 Implications from high-quality crystals

Figure 9(a) compares the $\rho$ data of two high-quality crystals with large RRR values of ~60 and low $\rho_0$ values of ~5 μΩ cm, which may be the highest and lowest ever reported, respectively. Although the two crystals possess similar qualities, a distinct anomaly in $\rho$ at $T_{s2}$ is observed only for crystal #40A: $\rho$ for crystal #40A shows a jump between 112 and 117 K with a negligible hysteresis, while only a smooth variation is observed for crystal #40B.

The jump in $\rho$ for crystal #40A is more pronounced than that for the previous crystal [#1A with RRR = 30; inset of Fig. 9(a)].[83] Moreover, the shape of the transition is different. First, the thermal hysteresis is smaller in crystal #40A, suggesting a lower density of defects. Second, the relation in the magnitudes of $\rho$ for phases II and III at the transition is reversed between crystals #40A and #1A: near the transition, $\rho$ for phase III is lower than that for phase II in #40A, and vice versa for #1A. As a result, the transitions appear as a jump and a plateau in crystals #40A and #1A, respectively. This fact implies that, as the sample quality improves, the resistivity or scattering of carriers is more effectively reduced in phase III than in phase II; in other words, the Fermi surface of phase III is more sensitive to impurity scattering. This could be one reason why the $T_{s2}$ transition in $\rho$ remained elusive: one would expect no change at $T_{s2}$ when only $\rho$ for phase III is enhanced by defects so as to accidentally match $\rho$ for phase II.

Another origin of the difference in $\rho$ between the two crystals is revealed by the heat capacity data shown in Fig. 9(b): upon cooling, a distinct peak is observed at 112 K for crystal #40A, while a broad hump is present at 95–115 K for crystal #40B. The shape of the peak for crystal #40A is characteristic of a first-order transition when measured by the relaxation method as in the present study. The broad hump for crystal #40B must be a result of the convolution of many peaks with different peak temperatures. Therefore, the $T_{s2}$ transition exists in both crystals but does not show up as a clear anomaly in $\rho$ for crystal #40B, because a cascade of transitions occur in a temperature range of 20 K. When this range becomes even larger, the hump in the heat capacity will be absorbed into the background, which may have been the case in most of the previous studies.[40,87]

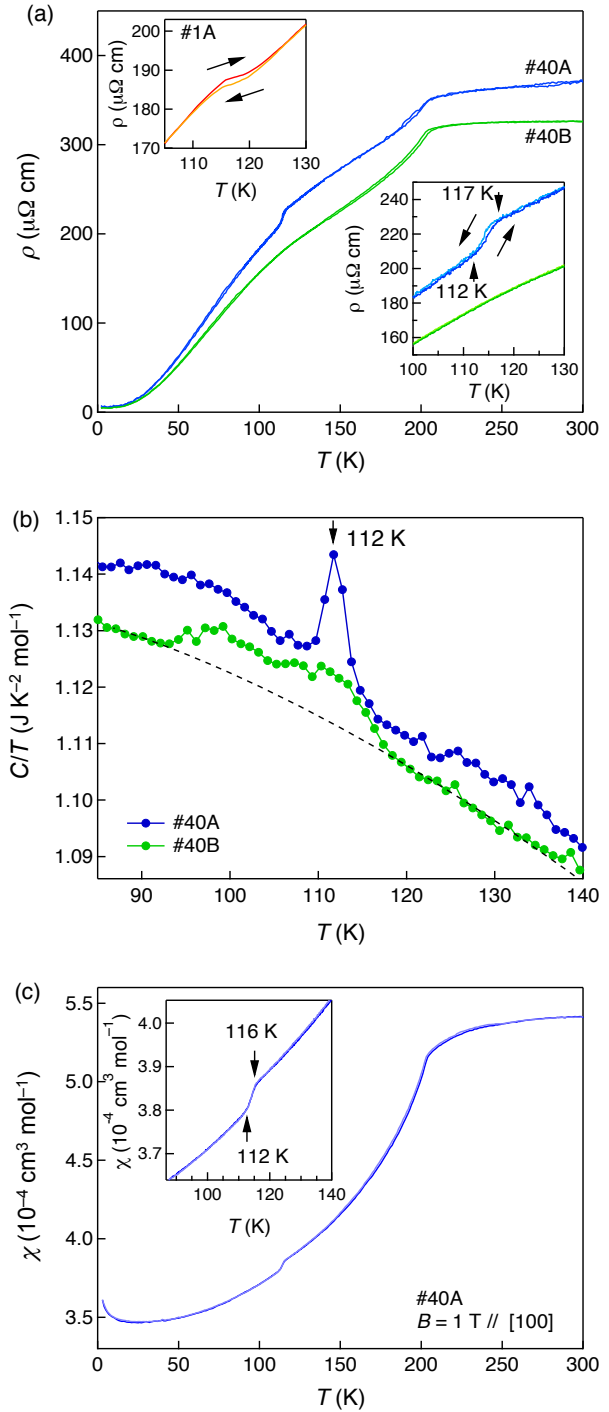

**Fig. 9.** (Color online) (a) Resistivity and (b) heat capacity of two typical high-quality crystals (#40A, #40B) with RRR ~ 60 and (c) magnetic susceptibility of crystal #40A measured at $B = 1$ T parallel to the [100] direction, showing a distinct anomaly at $T_{s2}$. The lower-right inset of (a) expands the $T$ range near $T_{s2}$, in which the cooling and heating curves are shown by the light-blue and blue lines, respectively. The upper-left inset of (a) shows the resistivity of the previous crystal #1A,[83] where the orange and red

curves represent the cooling and heating measurements, respectively. The broken line in (b) is an eye guide showing the possible background contribution.

The entropy changes $\Delta S$ associated with the $T_{s2}$ transition are estimated as 0.069 and 0.150 J K$^{-1}$ mol$^{-1}$ for the single peak for #40A and the broad hump for #40B, respectively, which are comparable to the values of 0.08–0.10 J K$^{-1}$ mol$^{-1}$ in previous reports.[45,83] The smaller $\Delta S$ value from the single peak for #40A suggests that only part of the sample contributed to the peak and there is also a distribution of the transition temperature even in this crystal; in fact, a broad hump seems to follow at lower temperatures of around 95 K in Fig. 9(b). It is plausible that a first-order transition with a minimal change in free energy is affected by a small number of defects or a tiny stress so that it occurs at different temperatures in different locations within a temperature range. This disorder effect may be more serious near the surface. In fact, surface-sensitive photoemission spectroscopy experiments revealed that the DOS decreases only by 5% below $T_{s1}$,[89] or it does not change at $T_{s1}$ but decreases below $T_{s2}$,[87,90] which are apparently inconsistent with many other experiments. Moreover, the absence of the $T_{s2}$ transition in the nonlinear optical measurements may be related to this surface problem.[28,36]

A novel observation using the high-quality crystal of #40A is a sudden change in magnetic susceptibility at $T_{s2}$, which has not been observed in the previous studies except for a tiny anomaly in the recent study by Tachibana et al.[45] As shown in Fig. 9(c), $\chi$ measured at $B = 1$ T along the [100] direction shows an apparent jump at 112–116 K, similar to the jump in $\rho$. The origin cannot be attributed to a change in the DOS, as in the case at $T_{s1}$, but must be related to a change in the Fermi surface at $T_{s2}$. The change in $\chi$ at $T_{s2}$ shows complex dependences on the field direction and strength, and also differences between zero-field- and field-cooling measurements, which will be reported elsewhere.

Finally, the transition temperatures of crystal #40A are 112–117, 112, and 112–116 K for the resistivity, heat capacity, and magnetic susceptibility, respectively; the previous crystal #1A showed a plateau at 116–119 K.[83] Because all the anomalies occur at 112 K on the low-temperature side, and because the sharp peak in the heat capacity at this temperature is the clearest signature of a phase transition, we take $T_{s2} = 112$ K for crystal #40A. This slightly lower value than the previous ones may be closer to the intrinsic critical temperature of the $T_{s2}$ transition in the clean limit.

## 7. Possible Orbital or Multipolar Fluctuations at AP

One of the most interesting implications from the previous NMR/NQR study is the presence of orbital fluctuations in Cd$_2$Re$_2$O$_7$.[18] $^{111}K_{iso}$ (and also $\chi_s$) and $^{111}(T_1T)^{-1}$ show similar $T$ dependences upon cooling (Fig. 7). However, in contrast to the saturation of $^{111}K_{iso}$ and $\chi_s$ at low temperatures, $^{111}(T_1T)^{-1}$ continues to decrease below ~50 K. This is probably related to the fact that $^{111}K_{iso}$ and $^{111}(T_1T)^{-1}$ can be used to probe the static susceptibility at the Brillouin zone center and the dynamical susceptibility averaged over the zone, respectively. Thus, there is additional suppression of the dynamical susceptibility at low temperatures.

It is important to note that the ratio of $T_1^{-1}$ at the $^{187}$Re and $^{111}$Cd sites is unexpectedly large: $^{187}T_1^{-1}/^{111}T_1^{-1} = 420$ at 5 K, which is much larger than the expected value of 17.[18] This means that there is an effective relaxation process that only occurs at the $^{187}$Re site, which is likely a fluctuation of 5$d$ orbital hyperfine fields. Very interestingly, $^{187}T_1^{-1}$ shows a steep increase above $T_{s2}$, in contrast to a smooth increase in $^{111}T_1^{-1}$, which suggests that the orbital fluctuations are even larger in phase II than in phase III.[85]

Concerning the mass enhancement, the experimental Sommerfeld coefficient $\gamma_{exp}$ is available only for phase III: 30.2 or 29.6 mJ K$^{-2}$ mol$^{-1}$.[1,3] On the other hand, the bare band mass $\gamma_{band}$ is not known for phase III and is 11.4 or 12.4 mJ K$^{-2}$ mol$^{-1}$ for phase I.[15,16] Provided that the DOS of phase I is twice as large as that of phase III, that is, $\gamma_{exp} \sim 60$ mJ K$^{-2}$ mol$^{-1}$, there could be a huge mass enhancement by a factor of 5 for phase I. For phase III, taking into account a possible band splitting from phase I, $\gamma_{band}$ may be smaller than that of phase I. Thus, the mass enhancement factor could be larger than 3. A moderate enhancement of the effective mass for phase III was inferred from optical spectroscopy.[91]

The large $\gamma_{exp}$ of ~ 30 mJ K$^{-2}$ mol$^{-1}$ and relatively small $\chi_s$ of $1.38 \times 10^{-4}$ cm$^3$ mol$^{-1}$ for phase III yield a small Wilson ratio of $R_W = 0.34$.[18,52] Such a small value of $R_W$ usually implies a strong $e$-$ph$ coupling; the importance of an $e$-$ph$ coupling for Cd$_2$Re$_2$O$_7$ has been evidenced by Raman scattering[42] and far-IR spectroscopy experiments.[70,91] On the other hand, for the related superconducting pyrochlore osmates AOs$_2$O$_6$, $R_W$ is also small and becomes even smaller with increasing electron-rattler interaction [0.48 (A = Cs), 0.36 (Rb), and 0.14 (K)];[75] the Cs and Rb compounds are weak-coupling superconductors, while the K compound is an extremely strong-coupling superconductor. Thus, the small $R_W$ values of Cd$_2$Re$_2$O$_7$ as well as the Cs and Rb osmates may not be due to $e$-$ph$ couplings. There may be an alternative route that selectively enhances $\gamma$ or reduces $\chi_s$.

The entity of the orbital fluctuation remains unclear. If one considers the importance of ASOC, it would more accurately be called a multipolar fluctuation, that is, a fluctuation associated with spin-orbital composite multipolar bands such as illustrated in Fig. 10. It is also important to note that there must be a secondary effect to induce a coupling to the lattice via ASOC to give an unconventional electron-lattice coupling. Possibly related to this, Hajialamdari and co-workers pointed out the strong presence of phonons in their optical reflectance spectrum, which is unusual in such a metal where conventional phonon modes should be strongly damped and barely observable.[70]

The multipolar fluctuation should be further enhanced approaching the QCP of the ISB transition under high pressure in $Cd_2Re_2O_7$, where one would expect a prominent influence of the fluctuation on the superconducting as well as the normal-state properties. The large enhancement in the coefficient of the $T^2$ resistivity near $P_c$ must be related to this. Therefore, the multipolar fluctuation serves as a strong scattering source for the electron transport in $Cd_2Re_2O_7$.

## 8. High-Pressure Properties
### 8.1 Theoretical implications

A key feature theoretically suggested for the SOCM by Fu and coworkers[24,26] and Wang et al.[27] is schematically depicted in the conceptual phase diagram in Fig. 10. Upon cooling below the ISB line, an SOCM is transformed into an itinerant multipolar state with spin-split bands induced by ASOC. As a function of a certain control parameter, the ISB line becomes lower and eventually vanishes at a QCP. Near the critical point on the left side of the ISB line, fluctuations associated with the multipolar order can induce $s$-wave-dominant and $p$-wave-dominant superconducting states, which are separated by a first-order transition line.[27] On the right of the ISB line, on the other hand, the multipolar fluctuations can cause purely $p$-wave superconductivity; the simplest case of the odd-parity pairing channel is expected to be a full gap state with time-reversal symmetry, as in the Balian–Werthamer B-phase of liquid $^3$He.[92]

Dresselhaus-type ASOC). As a function of a certain control parameter such as pressure, the ISB line can be suppressed and vanishes at a quantum critical point, where a superconducting dome composed of three regimes is expected: $s$-wave-dominant and $p$-wave-dominant regions, and a region with only $p$-wave pairings.[27]

### 8.2 $Cd_2Re_2O_7$ under high pressures

It would be fascinating, if this scenario really applies to $Cd_2Re_2O_7$. In the case of $Cd_2Re_2O_7$, the control parameter can be hydrostatic pressure. Early HP studies found that the ISB line disappears between 2.8 and 5.3 GPa and that both the coefficient of the $T^2$ resistivity and the residual resistivity are enlarged at 3.5–4 GPa.[64,93] Moreover, $T_c$ increases to 2 K at 2 GPa. Later, Kobayashi et al. carried out precise resistivity measurements under probably more hydrostatic pressures up to 4.6 GPa and observed much strange pressure dependences, as reproduced in Fig. 11:[65] the $T_{s1}$ line tends to vanish at $P_c \sim 4.2$ GPa, where five electronic phases, which are separated from each other by first-order transition lines, appear in addition to the three phases at AP. The $T_{s2}$ transition could not be traced in their study, because the anomaly in the resistivity was weak at AP and became indiscernible under HPs. Very recently, Kobayashi et al. carried out another series of experiments using a high-quality crystal with RRR ~ 50 and observed a gradual decrease in $T_{s2}$, as shown in Fig. 11, which will be reported elsewhere. This $T_{s2}$ line coincides with that obtained by Barišić et al. in their thermopower measurements,[4] which is also plotted in Fig. 11.

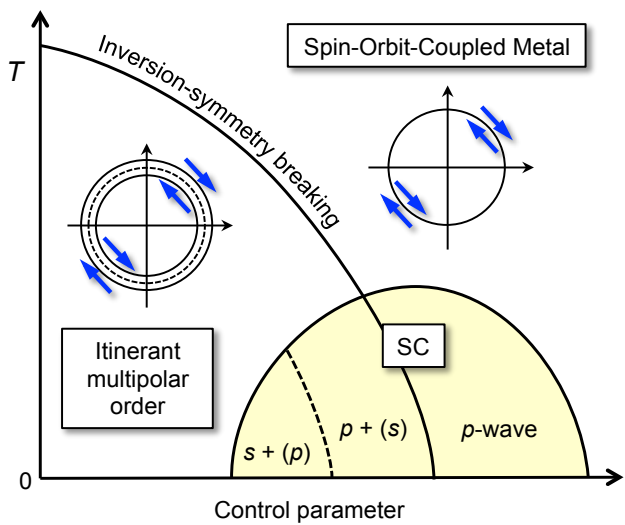

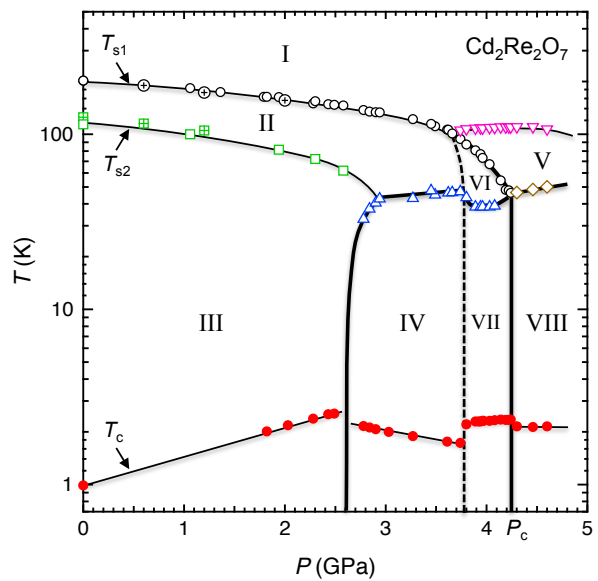

**Fig. 10.** (Color online) Schematic phase diagram for a spin-orbit-coupled metal (SOCM).[26,27] Owing to the inherent Fermi liquid instability of an SOCM, an itinerant multipolar order can emerge accompanied by an inversion-symmetry-breaking (ISB) transition, where an antisymmetric spin-orbit coupling (ASOC) is activated, resulting in spin-split bands such as illustrated (a Rashba-type spin splitting in a two-dimensional Fermi surface is assumed here, but a more complex splitting is expected for $Cd_2Re_2O_7$ due to a

**Fig. 11.** (Color online) $P$–$T$ phase diagram of $Cd_2Re_2O_7$.[65] The open marks with and without crosses represent phase transitions determined by thermopower[4] and resistivity measurements, respectively.[65] The green open squares on the $T_{s2}$ line are new data obtained by resistivity measurements using a high-quality crystal of RRR ~ 50. The red filled circles represent the values of $T_c$. All the lines are eye guides to show possible phase boundaries. The thin and thick lines indicate second- and first-order transitions, respectively. The nearly vertical broken line at ∼ 3.8 GPa was determined by the resistivity measurements but was not observed in the XRD study.[23] Suggested monoclinic space groups for the HP phases are $Cm$ (phase VI), $Cc$ (IV, VII), $C2/c$ (V), and $C2/m$ (VIII).

Concerning the superconductivity, $T_c$ increases linearly from 1.0 to 2.5 K in phase III, and suddenly drops at the III–IV boundary at 2.6 GPa, followed by a linear decrease in phase IV. Then, there are two small drops at both ends of the almost flat $T_c$ curve in phase VII. Thus, the four LT phases are clearly distinguished by the subtle differences in $T_c$ as well as its pressure dependences. $B_{c2}$ also increases markedly from 0.29 T at AP to 1.8 T (2 GPa), 3.2 T (3.3 GPa), and 7.8 T (4 GPa), which is 27 times larger than at AP and oversteps the Pauli-limiting field of 4 T. Then, it decreases slightly to 6.2 T at 4.5 GPa above $P_c$.[65]

*8.3 Multiple multipolar phases?*

Yamaura et al. have examined the crystal structures of $Cd_2Re_2O_7$ under HPs by high-resolution synchrotron powder XRD.[23] They revealed that the vanishing $T_{s1}$ line from the resistivity measurements in fact corresponds to the ISB line: the 002 reflection of the cubic cell, which is forbidden for centrosymmetric structures, appears only below the $T_{s1}$ line. More recently, Raman scattering experiments at 15 K found the recovery of inversion symmetry above $P_c$.[94] Therefore, the theoretical phase diagram in Fig. 10 seems to be realized to some extent in $Cd_2Re_2O_7$: AP may correspond to the left edge of the superconducting dome and the QCP occurs at $P_c$. Hence, the enhancement of $B_{c2}$ must be related to the enhancement of the $p$-wave channel near the QCP.

It has also been shown that most of the electronic phases take different crystal structures.[23] Very small distortions to monoclinic or triclinic unit cells are observed for phases IV–VIII. Taking into account of the group-subgroup relation, possible monoclinic space groups are $Cm$ (phase VI), $Cc$ (IV, VII), $C2/c$ (V), and $C2/m$ (VIII). It is noted, however, that there are two inconsistencies between the resistivity and XRD results.[23,65] One is that the latter XRD study shows a monoclinic distortion below ∼120 K inside the ISB dome instead of the $T_{s2}$ line above phase III ($I4_122$) in Fig. 11; no change is detected at the $T_{s2}$ line and a new monoclinic phase IX in $Cm$ is suggested. The other discrepancy is that the XRD study fails to find differences between phases IX (II) and VI and also between phases IV and VII; they are assigned to $Cm$ and $Cc$, respectively. Thus, the vertical line at 3.8 GPa does not seem to exist in terms of the crystal structure. If so, it might be possible for two-phase mixtures to exist in the regions of phases VI and VII; in fact, the large enhancement in residual resistivity[65] can be ascribed to scattering due to the two-phase mixture, as in the case of alloys.[95] Alternatively, these results would mean that there are electronic transitions without changes in the crystal structure at 3.8 GPa; the first-order character in the resistivity data may be compatible with this fact. These two controversial issues on the $P$–$T$ phase diagram of $Cd_2Re_2O_7$ should be clarified in future study.

HP measurements by Malavi et al. found a transition to the $R$–$3m$ phase with inversion symmetry above 21 GPa at room temperature, which may be a nonmetallic phase.[96] The occurrence of so many HP phases indicates intimate coupling between the electronic and structural instabilities. It is likely that there are various ways to resolve the Fermi liquid instability of the SOCM in $Cd_2Re_2O_7$. In fact, Fu pointed out the possibility of another order parameter for multipolar phases instead of the two-dimensional $E_u$ representation at AP, that is, the three-dimensional $T_{2u}$ representation;[24] this order parameter was claimed by Harter et al. for the $T_{s1}$ transition at AP.[28] Note that the observed $Cm$ and $Cc$ phases with odd parity are not induced by the $E_u$ instability but by other instabilities such as $T_{1u}$ and $T_{2u}$.[31] It is also pointed out that there is another instability toward the $C2/c$ and $C2/m$ structures with even parity. As a result, the experimental phase diagram of $Cd_2Re_2O_7$ appears richer and more complex than the theoretical phase diagram.[23,65] It will be important to investigate the characters of these phase transitions under high pressures and the nature of superconductivity near $P_c$ in future.

## 9. Conclusion

We have revisited the α-pyrochlore oxide superconductor $Cd_2Re_2O_7$ from the materials point of view. The following conclusions are given:

1. The intrinsic, clean-limit $T_c$ of $Cd_2Re_2O_7$ is 0.97 K. Disorder effects cause an increase in $T_c$ in the bulk as well as at the crystal surface. The heat capacity of a high-quality crystal shows a slight deviation from the BCS form, suggesting a possible effect of parity mixing for the noncentrosymmetric superconductor.
2. $T^2$ resistivity is not dominant at AP; the $T^2$ contribution with a small coefficient originates from the Koshino–Taylor term. Thus, electron correlations play a minor role in this compound. In contrast, the $T^2$ resistivity becomes dominant under HPs, which might be due to orbital or multipolar fluctuations.
3. The $T_{s2}$ transition manifests itself in the resistivity, heat capacity, and magnetic susceptibility in high-quality crystals. A subtle but significant change in the Fermi surface may take place at the transition.

In addition, we have addressed our present understanding and future prospects regarding the crystal and electronic structures, possible orbital fluctuations, and high-pressure properties of $Cd_2Re_2O_7$. We hope that the physics of the spin-orbit-coupled metals will be enriched through future experimental and theoretical studies on $Cd_2Re_2O_7$.

**Acknowledgments**

This work was partly supported by the Core-to-Core Program for Advanced Research Networks given by Japan Society for the Promotion of Science. We thank H. Harima for providing the DOS data shown in Fig. 2. We also thank M. Takigawa, T. Hasegawa, H. Harima, and M. Dmitrii for the helpful discussion. The resistivity of crystal #10A shown in Fig. 6 was measured by Y. Irie, and the heat capacity of crystal #41A in Fig. 3(a) was measured by Y. Haraguchi.

*hiroi@issp.u-tokyo.ac.jp